\documentclass[doublecol]{epl2} 
\usepackage{graphicx}
\usepackage{dcolumn}
\usepackage{bm}
\usepackage{amssymb}
\usepackage{amsmath}
\begin{document}
\setcounter{page}{1}
\vskip 2cm
\title{Non linear massive gravity as a gravitational $\sigma$-model}
\shorttitle{\textbf{Non linear massive gravity as a gravitational $\sigma$-model}}

\author{Ivan Arraut \inst{1} \and Kaddour Chelabi \inst{2}}
\shortauthor{Ivan Arraut \etal}
\institute{                    
  \inst{1} Department of Physics, Faculty of Science, Tokyo University of Science,
1-3, Kagurazaka, Shinjuku-ku, Tokyo 162-8601, Japan\\
  \inst{2} State Key Laboratory of Theoretical Physics, Institute of Theoretical Physics, Chinese Academy of Science, Beijing 100190, Peoples Republic of China.}

\pacs{11.15.Ex}{Spontaneous breaking of gauge symmetries}
\pacs{14.70.Kv}{Gravitons}
\pacs{11.15.-q}{Gauge field theories}

\abstract{We show the direct analogy between the ghost-free non-linear formulation of massive gravity and the standard $\sigma$-models well understood in the literature. This issue explains why there are two non-trivial family of solutions for the spherically symmetric case inside the non-linear massive gravity formulations with two free-parameters $\alpha$ and $\beta$. In general, the case $\beta<\alpha^2$ has a single physical vacuum state. On the other hand, the case $\beta=\alpha^2$ contains a natural vacuum degeneracy. This is in perfect analogy with the $\sigma$-model for scalar fields where depending on the values taken by the parameters of the theory the vacuum can be single or degenerate.}

\maketitle
 
\maketitle 
\section{Introduction}
Massive gravity theories reproduce the idea of a massive graviton in order to explain why gravity becomes weaker at large scales such that the accelerated expansion of the universe takes place \cite{Mass2}. The most popular approach for massive gravity theories is the de-Rham-Gabadadze-Tolley (dRGT) formulation \cite{deRham}. Not only because it is ghost-free. But also for being able to reproduce naturally the appropriate value of the Vainshtein radius which in this theory (in other theories too) is the scale after which the accelerated expansion of the universe becomes relevant \cite{My paper1}. This scale is in perfect analogy with the one obtained inside General Relativity (GR) for the Schwarzschild de-Sitter case (S-dS), where the gravitational attractive effects of gravity and the repulsive effect of the cosmological constant ($\Lambda$) cancel each other \cite{M}. Inside the scenario of massive gravity, some black-hole solutions have been derived \cite{Kodama, BHS}. Among all the proposals, the most generic and the easiest one to analyze, is obtained in \cite{Kodama}. What is remarkable about the solution formulated in \cite{Kodama} is that it condensates two family of solutions in a single one. In fact, by defining the two free-parameters of the theory as $\alpha$ and $\beta$, the two family of solutions can be classified in {\bf 1). Type I:} Which corresponds to the condition $\beta<\alpha^2$. {\bf 2). Type II:} Which corresponds to the condition $\beta=\alpha^2$. Initially, it would seem to be a coincidence the existence of precisely two family of solutions depending on the parameter combination. However, after some deep analysis, it is possible to demonstrate that there is a perfect correspondence between the two families and the notions of vacuum. In fact, an alternative classification is {\bf 1). Type I:} Corresponding to a unique vacuum in a free-falling frame. {\bf 2). Type II:} Corresponding to a degenerate vacuum in a free-falling frame \cite{Higgs}. For simplicity, it is possible to impose the stationary condition. In any case, in general it is always expected that any deviation with respect to the stationary condition should come from the perturbations of the metric. If gravity is included, the two type of solutions can be classified as follows {\bf 1). Type I:} The vacuum is uniquely defined by the fundamental scales of the theory and the location of the observer with respect to the source. {\bf 1). Type II:} The vacuum is degenerate and then the fundamental scales of the theory are arbitrary \cite{Higgs, Deg}. Similar results were found inside bigravity formulations in \cite{BabiandBrito} but analyzed from a different perspective.           

\section{The scalar $\sigma$-model}   \label{sigmamodel}

The scalar $\sigma$-model is defined by the Lagrangian \cite{Peskin}

\begin{equation}   \label{eq:Thelag}
\pounds=\frac{1}{2}(\partial_\mu\phi^i)^2+\frac{1}{2}\mu^2(\phi^i)^2-\frac{\lambda}{4}[(\phi^i)^2]^2,
\end{equation}
containing the two free-parameters $\lambda$ and $\mu$. Depending on the value taken by $\mu^2$ (or the relation between the two free-parameters), the vacuum is unique or degenerate. Note that the previous action is invariant under the group of transformations

\begin{equation}   \label{phila} 
\phi^i\to R^{ij}\phi^j.
\end{equation}
Here we will represent this transformation schematically $M=O(N)$ in order to represent the set of orthogonal matrices defined by eq. (\ref{phila}). The potential is defined by 
  
\begin{equation}
V(\phi^i)=-\frac{1}{2}\mu^2(\phi^i)^2+\frac{\lambda}{4}[(\phi^i)^2]^2.
\end{equation}
The potential has a minimum defined by the condition

\begin{equation}
\frac{\partial V(\phi^i)}{\partial \phi^i}=0. 
\end{equation}
Note that if $\mu^2<0$, the minimum is located at $\phi^i=0$. On the other hand, if $\mu^2>0$, then the vacuum is degenerate and located at 

\begin{equation}   \label{eq:Iseelanola}
(\phi^i_0)^2=\frac{\mu^2}{\lambda}.
\end{equation}
This minimum is degenerate and then we can select with arbitrariness the direction of the vacuum field configuration. Normally, the vacuum direction is taken such that 

\begin{equation}
\phi^i_0=(0,0,...,0,v),\;\;\;\;\;with\;\;\;\;\; v=\mu/\sqrt{\lambda}.
\end{equation} 
Note that the vacuum is invariant under the subgroup $H=O(N-1)$, which leaves the selected axis unchanged. The subgroup only exchange massless particles among themselves. The dimension of the coset space, defined as $M/H$, provides the number of broken generators. The review of these results can be found in \cite{Peskin}.  

\section{Massive gravity black-hole solutions}   \label{MBHSLL}

The spherically symmetric black-hole in dRGT massive gravity, contains two non-trivial solutions (not conformal to Minkowski) \cite{Kodama}. The action in massive gravity can be expressed schematically as

\begin{equation}
S=(Kinetic)-(Potential),
\end{equation}
where the Kinetic term corresponds to the Einstein-Hilbert action after expanding it up to second order, in other words, it contains second-order derivative terms in the action as follows $\sqrt{-g}R\to (\partial h)(\partial h)$. The potential is a term containing non-derivative terms interactions in the action. In gravity, the field is the graviton and it is represented by the perturbations of the metric $g_{\mu\nu}=g_{\mu\nu back}+h_{\mu\nu}$, with $g_{\mu\nu back}$ representing the background solution. The explicit expression for the action in massive gravity is 

\begin{equation}   \label{eq:b1}
S=\frac{1}{2\kappa^2}\int d^4x\sqrt{-g}(R+m_g^2U(g,\phi)),
\end{equation}
where $U(g,\phi)$ is equivalent to

\begin{equation}   \label{eq:Valentine}
U(g, \phi)=U_2+\alpha_3U_3+\alpha_4U_4. 
\end{equation}
The exact definitions of $U_n$, can be found in \cite{deRham, Kodama}. As can be observed from the definition (\ref{eq:Valentine}), the potential in massive gravity, also contains two free-parameters as it is the case of the linear $\sigma$-model. It is convenient to re-define the set of parameters by 

\begin{equation}
\alpha=1+\alpha_3,\;\;\;\;\;\beta=3(\alpha_3+4\alpha_4).
\end{equation}
With this definition, the Schwarzschild de-Sitter solutions in massive gravity theories, develop the following results:

\begin{equation}
ds^2=-f(Sr)dT_0^2(r,t)+\frac{S^2}{f(Sr)}dr^2+S^2r^2d\Omega^2,
\end{equation}
as it was demonstrated in \cite{Kodama}. This generic solution can be classified by the following cases

\subsection{Type I: $\beta<\alpha^2$}

In a free-falling frame, where we can define $f(Sr)\to1$, the vacuum for this case is unique. This means that the minimal of the potential defined by 

\begin{equation}   \label{eq:vac}
\frac{\partial V}{\partial h_{\mu\nu}}=0,
\end{equation}
where $V(g,\phi)=\sqrt{-g}U(g,\phi)$ is unique. In other words, the solution for the previous equation is $h_{\mu\nu vac}=h_{\mu\nu}(\alpha^2, \beta)$ trivially for the free-falling case. Note that this is connected to the fact that $T_0'(r,t)=0$ under the free-falling condition previously mentioned for this case in \cite{Kodama, Higgs, Deg}. Once the parameters of the theory are fixed, then the vacuum is automatically defined. The vacuum defined in this way is invariant under diffeomorphism transformations and as a consequence, the symmetry is not spontaneously broken in this case. Note that the free-falling condition eliminates the fundamental scales of the theory, namely, the Newtonian constant ($G$) and the graviton mass parameter ($m$).    

\subsection{Type II: $\beta=\alpha^2$}

For this second family of solutions, for the free-falling condition, the vacuum is degenerate in the sense that the solution for the condition (\ref{eq:vac}) is given by 

\begin{equation}   \label{eq:Momoko1}
h_{\mu\nu vac}=h_{\mu\nu}(\alpha, T_0'(r,t)),
\end{equation}
with $T_0'(r,t)$ arbitrary. Note that here we are assuming the stationary condition $\dot{T}_0(r,t)=S$. However, this condition is not necessary when we consider perturbations. Here we keep it for simplicity but in general, the vacuum solution as it is formulated in eq. (\ref{eq:Momoko1}), depends on $\dot{T}_0(r,t)$ too. There is no loss of generality in the stationary condition assumption because all our arguments related to $T_0'(r,t)$ can be just repeated for the case when there are deviations with respect to the stationary condition under the general solution $T_0(r,t)\approx St+A^t(r,t)$, as far as the conditions $A^t<<1$ and $\dot{A}^t<<1$ are satisfied \cite{Kodama, Higgs, Deg}. Note that the vacuum defined by eq. (\ref{eq:Momoko1}) is degenerate and is not invariant under the diffeomorphism transformations depending explicitly on the St\"uckelberg functions $T_0(r,t)$. This issue can be observed better if we define the dynamical metric containing all the degrees of freedom using the non-linear St\"uckelberg trick \cite{K}

\begin{equation}   \label{eq:stuefashio}
g_{\mu\nu}=\left(\frac{\partial Y^\alpha}{\partial x^\mu}\right)\left(\frac{\partial Y^\beta}{\partial x^\nu}\right)g'_{\alpha\beta},
\end{equation}
with $Y^0(r,t)=T_0(r,t)$ and $Y^r=Sr$ as has been demonstrated in \cite{Kodama, Higgs}. The diffeomorphism transformations are then defined by

\begin{equation}   \label{eq:gt}
g_{\mu\nu}\to\frac{\partial f^\alpha}{\partial x^\mu}\frac{\partial f^\beta}{\partial x^\nu}g_{\alpha\beta}(f(x)), \;\;\;\;\;Y^\mu(x)\to f^{-1}(Y(x))^\mu.
\end{equation}
Note that the vacuum definition (\ref{eq:Momoko1}) will not be invariant under the set of transformations in eq. (\ref{eq:gt}) depending explicitly on the St\"uckelberg function $Y^\alpha(x)$. In this case then we say that the symmetry is spontaneously broken. The broken generators correspond to the set of transformations containing the explicit dependence on $T_0(r,t)$. In this sense, the vacuum is clearly not unique. The classification of the solutions in {\bf Type I} and {\bf Type II} in connection with what happens with the ordinary sigma models, is one of the new results of this paper. Note that once again the free-falling condition eliminates the fundamental scales of the theory locally.  
     
\section{Including gravity}   \label{inclduding la}

When gravity is included, we have to select the location of the observer with respect to the source. In such a case, we can generalize the result such that $f(Sr)\approx 1-\epsilon$, with $\epsilon$ containing the fundamental scales of the theory, namely, $G$ and $m$. If we select the location of the observer with respect to the source using $r\backsim(GM/m^2)^{1/3}$, then $\epsilon=(GMm_g)^{2/3}$ \cite{Higgs, Deg}. However, it is not necessary to specify in principle any location. What is important for the moment is to keep in mind that $\epsilon$ contains the fundamental scales of the theory. Again here we have two type of solutions and they can be divided depending on the relation between the two free-parameters as follows

\subsection{Type I: $\beta<\alpha^2$}

For which the St\"uckelberg function has to satisfy the constraint \cite{Kodama}

\begin{equation}   \label{eq:wonderful}
(T_0'(r,t))^2=\frac{S^2(1-f(Sr))}{f(Sr)}\left(\frac{1}{f(Sr)}-1\right).
\end{equation}
If we introduce the result $f(Sr)\approx 1-\epsilon$, then the previous result can be expanded as follows

\begin{equation}
\vert T_0'(r,t)\vert \approx S \sum_{n=1}^{\infty}\epsilon^n,
\end{equation}
for $\epsilon <<1$. Or

\begin{equation}
\vert T_0'(r,t)\vert\approx -S \sum_{n=0}^{\infty}\epsilon^n,
\end{equation}
for $\epsilon >>1$. Here however we do not consider the case $\epsilon>>1$ because it would correspond to the situation where the observer is located inside the event horizon and the coordinates used for describing the solution become inappropriate. The case $\epsilon\to1$ corresponds to the event horizon condition. For the situation with $\epsilon<<1$, the vacuum solutions defined by eq. (\ref{eq:vac}) become

\begin{equation}
h_{\mu\nu vac}=h_{\mu\nu} (\alpha, \beta, \epsilon). 
\end{equation} 
The vacuum is then uniquely defined by the location of the observer with respect to the source and by the fundamental scales of the theory. 

\subsection{Type II: $\beta=\alpha^2$} 

For this situation, the constraint (\ref{eq:wonderful}) is not satisfied. In such a case, the function $T_0(r,t)$ is completely arbitrary. The arbitrariness of the function is then reflected on the arbitrariness of the parameter $\epsilon$ defined previously. By continuity in the solutions with respect to the parameters $\alpha$ and $\beta$, still we can assume a polynomial expansion in terms of the fundamental scales of the theory, or equivalently, in terms of $\epsilon$. Then we can define at the lowest order \cite{Higgs, Deg}

\begin{equation} 
T_0'(r,t)\backsim \epsilon,
\end{equation}
with $\epsilon$ arbitrary. Then the arbitrariness of the function $T_0(r,t)$ is equivalent to an arbitrariness in the fundamental scales of the theory, contained inside $\epsilon$. This issue might be relevant at the moment of locating fixed points inside this theory. The classification of the black-hole solutions in {\bf Type I} and {\bf Type II}, together with the related behavior of the fundamental constants of the theory, is the main contribution of this paper. The following tables illustrates the analogy between the linear $\sigma$-model and the non-linear formulation of massive gravity.  

\begin{center}
    \begin{tabular}{| l | l | l | l | }
    \hline
    {\bf \;\;\;\;\;Theory} & {\bf \;Kinetic term} & {\bf \;Potential} & {\bf Parameters}\\ \hline
    Massive gravity & $\sqrt{-g}R\to\partial h\partial h$ & $\sqrt{-g}U(g,\phi)$ & \;\;\;\;\;\;\;\;$\alpha^2 ,\beta$\\ \hline
    $\;\;\;\;\sigma$-model & \;\;\;\;\;\;\;$(\partial_\mu\phi^i)^2$ &\;\;\;\; $V(\phi^i)$ & \;\;\;\;\;\;\;\;$\mu^2, \lambda.$\\ 
   \hline
    \end{tabular}
\end{center}

\begin{center}
    \begin{tabular}{| l | l | l |}
    \hline
    {\bf \;\;\;\;\;Theory} & {\bf Unique Vac.} & {\bf Degenerate Vac.} \\ \hline
    Massive gravity & \;\;\;\;\;\;\;$\beta<\alpha^2$ & \;\;\;\;\;\;\;$\beta=\alpha^2$\\ \hline
    $\;\;\;\;\sigma$-model & \;\;\;\;\;\;\;$\mu^2<0$ & \;\;\;\;\;\;\;$\mu^2>0$ \\ 
   \hline
    \end{tabular}
\end{center}
The analogy suggests that Massive gravity theories can be observed as a special case of sigma models. In the coming section, we will review some other important results derived recently by other authors and pointing to this direction. The method proposed here analyze the vacuum solutions of the theory under spherical symmetry and then classify them in analogy with the sigma models. The analysis of the coming section, is based on the study of the mathematical structure of the massive action itself ($\sqrt{-g}U(g,\phi)$) and its connection with the generic non-linear sigma models.

\section{Non-linear sigma models and massive gravity}   \label{pla}

It has been proposed previously in \cite{de Rham}, that massive gravity formulations can be understood as a special case of the non-linear sigma models. The action for these type of models is defined by

\begin{equation}
S_{NLSM}=-\int d^Dx\frac{1}{2}\eta^{\mu\nu}\partial_\mu\phi^a\partial_\nu\phi^bf_{ab}(\phi).
\end{equation}
This generic form of a non-linear sigma model maps a base manifold toward a target space. The base manifold is usually taken to be Minkowski and the target space $f_{ab}(\phi)$ is usually required to be positive definite for avoiding ghost. This means that the target space has to be Riemannian. If the target space is symmetric, this statement is equivalent to the isometry group being compact. This condition is dubbed as the Riemann/compact requirement. However, there are exceptional cases where the target manifold is not required to be positive definite. One example corresponds to the Nambu-Goto action \cite{de Rham, Schwarz}. In such a case, although the target space is Lorentzian (pseudoRiemannian), the ghost can be avoided because it corresponds to a degree of freedom that can be gauged away through a first class constraint. Another way for avoiding ghosts in pseudoRiemannian target spaces is by introducing subgroups of gauge symmetries in the action such that the ghost is again projected out. Examples of these cases are given in supergravity constructions \cite{Super}. These approaches eliminate the ghost by using first class constraints. In general, given a non-compact group $G$, such that the Killing metric contains both, positive and negative signatures, it is possible to avoid the ghost by introducing a subgroup of gauge symmetries such that all the positive or all the negative signatures are projected out. On the other hand, by using second class constraints, it has been demonstrated that there is an unique way for avoiding the Riemann/compact condition for a target space with one negative direction (Lorentzian). The models where this can be done are defined by the action \cite{de Rham}

\begin{equation}   \label{eq:Alejandromario}
S_{mg\sigma}=\int d^Dx\sum^D_{n=1}\beta_n(\phi)X^{\mu_1}_{[\mu_1}X^{\mu_2}_{\mu_2}...X^{\mu_n}_{\mu_n]}.
\end{equation}       
Here $\beta_n(\phi)$ are functions of the St\"uckelberg fields $\phi^a$ and the metric $f_{ab}$ has a Lorentzian signature. The matrices $X^{\mu}_{\nu}$ are defined by

\begin{equation}
X^{\mu}_{\nu}=\sqrt{\eta^{\nu\rho}\partial_\rho\phi^a\partial_\nu\phi^bf_{ab}(\phi)},
\end{equation}
with $a=0,1,...,N-1$ and $N\geq D$. The potential terms in the non-linear massive gravity formulations, here defined by $\sqrt{-g}U(g,\phi)$ in eqns. (\ref{eq:b1}) and (\ref{eq:Valentine}) can be considered as a limit or special case of the result (\ref{eq:Alejandromario}) when the functions $\beta_n(\phi)$ become constants, the target space becomes Minkowsni $f_{ab}=\eta_{ab}$ and $N=D$. The consistency of this result is the purpose of the analysis done in \cite{de Rham}. The fact that massive gravity formulations are equivalent to special cases of non-linear sigma models implies the possibility of having inside the solutions of the theory some of the properties of the sigma models. In this paper for example, we have demonstrated that there are two non-trivial family of black-hole solutions in massive gravity. Here we called them $\bf Type \;I$ with $\beta<\alpha^2$, where we have a single vacuum and $\bf Type \;II$ with $\beta=\alpha^2$, where the vacuum is degenerate. As can be observed, the type of solution depends on the relation between the two free-parameters of the theory. This is in agreement with the behavior of a sigma model where depending on the relation between the free parameters of the theory the vacuum can be single or degenerate. Note that here we do not consider the case $\beta>\alpha^2$ because it is unphysical. In fact, it represents the branch of solutions for which the scale factor $S$ becomes complex and then the dynamical metric will not be well defined.       

\section{Conclusions}   \label{conc}

In this paper we have demonstrated the analogy between massive gravity theories and the scalar $\sigma$-models. This fact explains why there are two family of black hole solutions well defined in this theory \cite{Kodama}. The family for which $\beta<\alpha^2$, gives an unique notion of vacuum, defined in general by the position of the observer with respect to the source. On the other hand, the family for which $\beta=\alpha^2$, corresponds to the case of degenerate vacuum, where the symmetry is spontaneously broken. When gravity appears, the arbitrariness of the St\"uckelberg function $T_0(r,t)$ for this solution is equivalent to the arbitrariness of the fundamental scales of the theory $G$ and $m$, this observation and the analysis involved is the main contribution of this paper. From this perspective, the theory of massive gravity seems to develop a rich structure at the vacuum level. We have to remark that the ambig\"uity in the notions of vacuum in massive gravity theories is what create an (apparent) effect of extra-particle creation at large scales inside these theories even if the physics at the event horizon level never changes with respect to the GR case \cite{HawkingdRGT, Vainshtein paper}. Note that the case $\beta>\alpha^2$ is not considered as a physical one in agreement with the solutions found in \cite{Kodama} because for this case, the scale factor $S$ becomes complex. Other results in the literature have also demonstrate that massive gravity theories are in essence a special case of non-linear sigma models \cite{de Rham}. The analysis done in \cite{de Rham} is focused in the potential term of the massive action of the theory and its connection with the non-linear sigma models. In this paper we have taken an already well derived solution inside the scenario of the non-linear formulation of massive gravity and we have analyzed the behavior of the vacuum solutions. The solution is then classified in two families depending on the relation between the two free-parameters of the theory. The results clearly suggests a typical behavior of a sigma model. We can then conclude that any modification of gravity based on sigma models should develop well defined families of black-holes depending on the relation between the two free-parameters of the theory.       
\\\\

{\bf Acknowledgement}
I. A would like to thank Petr Horava for the kind invitation to the University of California Berkeley where there was the opportunity of having discussions about these results. I.A. would like to thank Misha Smolkin for organizing the seminar where part of these results together with the notions of graviton Higgs mechanism were introduced. I. A. would like to thank Raphael Bousso, Hitoshi Murayama and Ziqi Yan for useful discussions during the visit of I. A. to Berkeley. I. A. would like to thank Henry Tye for the invitation to Hong Kong University of Science (HKUST), where there was the opportunity of sharing these and previous related ideas. I. A. would like to thank Jan Hajer for organizing the seminar in HKUST and Toshifumi Noumi for useful comments and discussions. The authors would like to thank Shigeki Sugimoto for the opportunity of participating in the School on String and Fields organized at the Yukawa Institute for Theoretical Physics (YITP) from Feb. 29th-March 4th. The authors would like to thank David Poland, Francesco Benini, Romuald Janik and Tadashi Takayanagi for nice discussions during the School. I.A would like to thank Nobuyoshi Ohta for the kind invitation to Kinki University in Osaka and for organizing the seminar where some of these ideas were introduced and debated. I.A. appreciate the discussions with Akihiro Ishibashi during the seminar at Kinki University in Osaka. I. A. is supported by the JSPS Post-doctoral fellowship for oversea Resarchers. K. C. is supported by the CAS Twast Presidential program for Ph.D. students.

\end{document}